\documentstyle[twoside,psfig]{article}

\catcode`\@=11
\long\def\@makefntext#1{
\protect\noindent \hbox to 3.2pt {\hskip-.9pt
$^{{\eightrm\@thefnmark}}$\hfil}#1\hfill}       %CAN BE USED

\def\thefootnote{\fnsymbol{footnote}}
\def\@makefnmark{\hbox to 0pt{$^{\@thefnmark}$\hss}}    %ORIGINAL

\def\ps@myheadings{\let\@mkboth\@gobbletwo
\def\@oddhead{\hbox{}
\rightmark\hfil\eightrm\thepage}
\def\@oddfoot{}\def\@evenhead{\eightrm\thepage\hfil
\leftmark\hbox{}}\def\@evenfoot{}
\def\sectionmark##1{}\def\subsectionmark##1{}}

\oddsidemargin=\evensidemargin
\renewcommand{\thefootnote}{\fnsymbol{footnote}}

\newcounter{sectionc}\newcounter{subsectionc}\newcounter{subsubsectionc}
\renewcommand{\section}[1] {\vspace{12pt}\addtocounter{sectionc}{1}
\setcounter{subsectionc}{0}\setcounter{subsubsectionc}{0}\noindent
    {\tenbf\thesectionc. #1}\par\vspace{5pt}}
\renewcommand{\subsection}[1] {\vspace{12pt}\addtocounter{subsectionc}{1}
    \setcounter{subsubsectionc}{0}\noindent
    {\bf\thesectionc.\thesubsectionc. {\kern1pt \bfit #1}}\par\vspace{5pt}}
\renewcommand{\subsubsection}[1]
{\vspace{12pt}\addtocounter{subsubsectionc}{1}
    \noindent{\tenrm\thesectionc.\thesubsectionc.\thesubsubsectionc.
    {\kern1pt \tenit #1}}\par\vspace{5pt}}
\newcommand{\nonumsection}[1] {\vspace{12pt}\noindent{\tenbf #1}
    \par\vspace{5pt}}

\newcounter{appendixc}
\newcounter{subappendixc}[appendixc]
\newcounter{subsubappendixc}[subappendixc]
\renewcommand{\thesubappendixc}{\Alph{appendixc}.\arabic{subappendixc}}
\renewcommand{\thesubsubappendixc}
    {\Alph{appendixc}.\arabic{subappendixc}.\arabic{subsubappendixc}}

\renewcommand{\appendix}[1] {\vspace{12pt}
        \refstepcounter{appendixc}
        \setcounter{figure}{0}
        \setcounter{table}{0}
        \setcounter{lemma}{0}
        \setcounter{theorem}{0}
        \setcounter{corollary}{0}
        \setcounter{definition}{0}
        \setcounter{equation}{0}
        \renewcommand{\thefigure}{\Alph{appendixc}.\arabic{figure}}
        \renewcommand{\thetable}{\Alph{appendixc}.\arabic{table}}
        \renewcommand{\theappendixc}{\Alph{appendixc}}
        \renewcommand{\thelemma}{\Alph{appendixc}.\arabic{lemma}}
        \renewcommand{\thetheorem}{\Alph{appendixc}.\arabic{theorem}}
        \renewcommand{\thedefinition}{\Alph{appendixc}.\arabic{definition}}
        \renewcommand{\thecorollary}{\Alph{appendixc}.\arabic{corollary}}
        \renewcommand{\theequation}{\Alph{appendixc}.\arabic{equation}}
        \noindent{\tenbf Appendix \theappendixc #1}\par\vspace{5pt}}
\newcommand{\subappendix}[1] {\vspace{12pt}
        \refstepcounter{subappendixc}
        \noindent{\bf Appendix \thesubappendixc. {\kern1pt \bfit #1}}
    \par\vspace{5pt}}
\newcommand{\subsubappendix}[1] {\vspace{12pt}
        \refstepcounter{subsubappendixc}
        \noindent{\rm Appendix \thesubsubappendixc. {\kern1pt \tenit #1}}
    \par\vspace{5pt}}

\newcommand{\textlineskip}{\baselineskip=13pt}
\newcommand{\smalllineskip}{\baselineskip=10pt}

\def\eightcirc{
\begin{picture}(0,0)
\put(4.4,1.8){\circle{6.5}}
\end{picture}}
\def\eightcopyright{\eightcirc\kern2.7pt\hbox{\eightrm c}}

\newcommand{\copyrightheading}[1]
    {\vspace*{-2.5cm}\smalllineskip{\flushleft
    {\footnotesize International Journal of Modern Physics A #1}\\
    {\footnotesize $\eightcopyright$\, World Scientific Publishing
     Company}\\
     }}

\def\abstracts#1#2#3{{
    \centering{\begin{minipage}{4.5in}\footnotesize\baselineskip=10pt
    \parindent=0pt #1\par
    \parindent=15pt #2\par
    \parindent=15pt #3
    \end{minipage}}\par}}

\newcommand{\bibit}{\nineit}

\renewenvironment{thebibliography}[1]
    {\frenchspacing
     \ninerm\baselineskip=11pt
     \begin{list}{\arabic{enumi}.}
    {\usecounter{enumi}\setlength{\parsep}{0pt}
     \setlength{\leftmargin 12.7pt}{\rightmargin 0pt} %FOR 1--9 ITEMS
     \setlength{\itemsep}{0pt} \settowidth
    {\labelwidth}{#1.}\sloppy}}{\end{list}}

\newcounter{itemlistc}
\newcounter{romanlistc}
\newcounter{alphlistc}
\newcounter{arabiclistc}
\newenvironment{itemlist}
        {\setcounter{itemlistc}{0}
     \begin{list}{$\bullet$}
    {\usecounter{itemlistc}
     \setlength{\parsep}{0pt}
     \setlength{\itemsep}{0pt}}}{\end{list}}

\newenvironment{romanlist}
    {\setcounter{romanlistc}{0}
     \begin{list}{$($\roman{romanlistc}$)$}
    {\usecounter{romanlistc}
     \setlength{\parsep}{0pt}
     \setlength{\itemsep}{0pt}}}{\end{list}}

\newenvironment{alphlist}
    {\setcounter{alphlistc}{0}
     \begin{list}{$($\alph{alphlistc}$)$}
    {\usecounter{alphlistc}
     \setlength{\parsep}{0pt}
     \setlength{\itemsep}{0pt}}}{\end{list}}

\newcommand{\fcaption}[1]{
        \refstepcounter{figure}
        \setbox\@tempboxa = \hbox{\footnotesize Fig.~\thefigure. #1}
        \ifdim \wd\@tempboxa > 5in
           {\begin{center}
        \parbox{5in}{\footnotesize\smalllineskip Fig.~\thefigure. #1}
            \end{center}}
        \else
             {\begin{center}
             {\footnotesize Fig.~\thefigure. #1}
              \end{center}}
        \fi}

\newcommand{\tcaption}[1]{
        \refstepcounter{table}
        \setbox\@tempboxa = \hbox{\footnotesize Table~\thetable. #1}
        \ifdim \wd\@tempboxa > 5in
           {\begin{center}
        \parbox{5in}{\footnotesize\smalllineskip Table~\thetable. #1}
            \end{center}}
        \else
             {\begin{center}
             {\footnotesize Table~\thetable. #1}
              \end{center}}
        \fi}

\def\@citex[#1]#2{\if@filesw\immediate\write\@auxout
    {\string\citation{#2}}\fi
\def\@citea{}\@cite{\@for\@citeb:=#2\do
    {\@citea\def\@citea{,}\@ifundefined
    {b@\@citeb}{{\bf ?}\@warning
    {Citation `\@citeb' on page \thepage \space undefined}}
    {\csname b@\@citeb\endcsname}}}{#1}}

\newif\if@cghi
\def\cite{\@cghitrue\@ifnextchar [{\@tempswatrue
    \@citex}{\@tempswafalse\@citex[]}}
\def\citelow{\@cghifalse\@ifnextchar [{\@tempswatrue
    \@citex}{\@tempswafalse\@citex[]}}
\def\@cite#1#2{{$\null^{#1}$\if@tempswa\typeout
    {IJCGA warning: optional citation argument
    ignored: `#2'} \fi}}

\def\pmb#1{\setbox0=\hbox{#1}
    \kern-.025em\copy0\kern-\wd0
    \kern.05em\copy0\kern-\wd0
    \kern-.025em\raise.0433em\box0}

\def\fnm#1{$^{\mbox{\scriptsize #1}}$}
\def\fnt#1#2{\footnotetext{\kern-.3em
    {$^{\mbox{\scriptsize #1}}$}{#2}}}

\def\thefootnote{\fnsymbol{footnote}}
\def\@makefnmark{\hbox to 0pt{$^{\@thefnmark}$\hss}}    %ORIGINAL

\def\ps@myheadings{%
    \let\@oddfoot\@empty\let\@evenfoot\@empty
    \def\@evenhead{\slshape\leftmark\hfil}%       %EVEN PAGE
    \def\@oddhead{\hfil{\slshape\rightmark}}%     %ODD PAGE
    \let\@mkboth\@gobbletwo
    \let\sectionmark\@gobble
    \let\subsectionmark\@gobble
    }
\font\tenrm=cmr10
\font\tenit=cmti10
\font\tenbf=cmbx10
\font\bfit=cmbxti10 at 10pt
\font\ninerm=cmr9
\font\nineit=cmti9

\font\eightrm=cmr8

\def\qed{\hbox{${\vcenter{\vbox{            %HOLLOW SQUARE
   \hrule height 0.4pt\hbox{\vrule width 0.4pt height 6pt
   \kern5pt\vrule width 0.4pt}\hrule height 0.4pt}}}$}}

\renewcommand{\thefootnote}{\fnsymbol{footnote}}  %USE SYMBOLIC FOOTNOTE

\begin{document}
\setlength{\textheight}{7.7truein}  %for 2nd page onwards

\thispagestyle{empty}

\markboth{\protect{\footnotesize\it V.V. Nesterenko,
G. Lambiase \& G. Scarpetta}}
{\protect{\footnotesize\it Casimir Energy of a Dilute
Dielectric Ball at Zero and Finite Temperature}}

\normalsize\textlineskip

\setcounter{page}{1}

\copyrightheading{}     %{Vol.~0, No.~0 (2000) 000--000}

\vspace*{0.88truein}

%\fpage{1}
\centerline{\bf CASIMIR ENERGY OF A DILUTE DIELECTRIC BALL}
\vspace*{0.035truein} \centerline{\bf AT ZERO AND FINITE
TEMPERATURE } \vspace*{0.37truein} \centerline{\footnotesize V.V.\
NESTERENKO}

\baselineskip=12pt \centerline{\footnotesize\it Bogoliubov
Laboratory of Theoretical Physics, Joint Institute for Nuclear
Research} \baselineskip=10pt \centerline{\footnotesize\it Dubna,
141980, Russia}

\vspace*{10pt} \centerline{\footnotesize G. LAMBIASE and G.
SCARPETTA} \baselineskip=12pt \centerline{\footnotesize\it
Dipartimento di Scienze Fisiche E.R. Caianiello, Universit\'a
 di Salerno} \baselineskip=10pt
\centerline{\footnotesize\it  Baronissi (SA), 84081, Italy}
\centerline{\footnotesize\it INFN, Sezione di Napoli,  Napoli,
80126, Italy} \vspace*{0.225truein}

%\publisher{(receiveddate)}{(revised date)}

\vspace*{0.21truein} \abstracts{The basic results in calculations of
the thermodynamic functions of electromagnetic field in the
background of a dilute dielectric ball  at zero and finite
temperature are presented. Summation over the angular momentum values
is accomplished in a closed form by making use of the addition
theorem for the relevant Bessel functions. The behavior of the
thermodynamic characteristics in the low  and high temperature limits
is investigated. The $T^3$-term in the low temperature expansion of
the free energy is recovered (this  term has been lost in our
previous calculations).}{PACS numbers: 12.20.Ds, 03.70.+k,
78.60.Mq}{}

%\textlineskip          %) USE THIS MEASUREMENT WHEN THERE IS
%\vspace*{12pt}         %) NO SECTION HEADING

\vspace*{1pt}\textlineskip  %) USE THIS MEASUREMENT WHEN THERE IS
\section{Introduction}    %) A SECTION HEADING
\vspace*{-0.5pt}
\noindent
%\section{}
%\subsection{Producing the Hard Copy}\label{subsec:prod}
Calculation of the Casimir energy of a dielectric ball has a rather
long history starting 20 years ago.$^1$ However only recently the
final result was obtained for a dilute dielectric ball at
zero\cite{2,LSN} and finite\cite{NLS,Barton} temperature. Here we
summarize briefly the derivation of the Casimir energy of a dilute
dielectric ball by making use of the mode summation method and the
addition theorem for the Bessel functions instead of the uniform
asymptotic expansion for these functions.\cite{LSN,NLS}

\section{The Basic Steps of Calculations and the Results}
\noindent A  solid ball of radius $a$ placed in an unbounded
uniform medium is considered. The contour integration
technique$^3$ gives ultimately the following representation for the
Casimir energy of the ball
\begin{equation}
 \label{1}
  E=-\frac{1}{2\pi a}\sum_{l=1}^{\infty}(2l+1)\int_0^{y_0}dy\,
 y\,\frac{d}{dy}\ln\left[W_l^2(n_1y, n_2y)-\frac{\Delta n^2}{4}\,
 P_l^2(n_1y, n_2y)\right]\,,
\end{equation}
where
\begin{eqnarray}
  W_l(n_1y,
n_2y)&=&s_l(n_1y)e_l^{\prime}(n_2y)-s_l^{\prime}(n_1y)e_l(n_2y)\,,
  \nonumber \\
  P_l(n_1y,
n_2y)&=&s_l(n_1y)e_l^{\prime}(n_2y)+s_l^{\prime}(n_1y)e_l(n_2y)\,,
\nonumber
\end{eqnarray}
and $s_l(x)$, $e_l(x)$ are the modified Riccati-Bessel functions,
$n_1, n_2 $ are the refractive indices of the ball and of its
surroundings, $\Delta n= n_1-n_2$.

Analysis of divergences carried out in our paper$^3$ leads to the
following algorithm for calculating the vacuum energy (\ref{1}) in
the $\Delta n^2$-approximation. First, the $\Delta n^2$--contribution
should be found, which is given by the sum $\sum_lW_l^2$. Upon
changing its sign to the opposite one, we obtain the contribution
generated by $W_l^2$, when this function is in the argument of the
logarithm. The $P^2_l$-contribution into the vacuum energy is taken
into account by expansion of Eq. (\ref{1}) in terms of $\Delta n^2$.

 Applying the addition theorem for the Bessel
functions\cite{Klich}
\[
  \sum_{l=0}^{\infty}(2l+1)[s_l^{\prime}(\lambda
  r)e_l(\lambda\rho)]^2=\frac{1}{2r \rho}\int_{r-\rho}^{r+\rho}
  \left(\frac{1}{\lambda}\,\frac{\partial{\cal D}}{\partial r}
 \right)^2R\;dR
\]
with
\[
{\cal D} =\frac{\lambda r\rho}{R}\, e^{-\lambda R}, \quad
  R=\sqrt{r^2+\rho^2-2r\rho\cos\theta}
\]
one arrives at the result$^£$
\[
  E=\frac{23}{384}\frac{\Delta n^2}{\pi a}
  =\frac{23}{1536}\,\frac{(\varepsilon_1-
  \varepsilon_2)^2}{\pi a}{,} \quad  \varepsilon_i=n^2_i, \quad i=1,2\,{.}
\]

Extension to finite temperature $T$ is accomplished by substituting
the $y$-integra\-tion in (\ref{1}) by summation over the Matsubara
frequencies $ \omega_n=2\pi nT$. When considering the  low
temperature  behavior of the thermodynamic functions of a dielectric
ball   the term proportional to $T^3$ in our paper$^4$ was lost. It
was due to the following. We have  introduced the summation over the
Matsubara frequencies in Eq.\ (3.20) under the sign of the
$R$-integral. Here we show how to do this summation in a correct way.

In the $\Delta^2$-approximation the last term in Eq.\ (3.20) from
the article$^4$
\begin{equation}
\label{eq-2} \overline {U}_W(T)=2 T\Delta n^2 \sum_{n=0}^\infty
\!{}^{'}w^2_n\int_{\Delta n}^2\frac{e^{-2w_nR}}{R}\,dR{,} \quad
w_n=2 \pi na T
\end{equation}
can be represented in the following form
\begin{equation}
\label{eq-3}
\overline {U}_W(T)=-2 T\Delta n^2
\sum_{n=0}^\infty\!{}^{'} w^2_n \,E_1(4w_n){,}
\end{equation}
where $E_1(x)$ is the exponential-integral function.$^7$
Now we accomplish the summation over the Matsubara frequencies by
making use of the Abel-Plana formula
\begin{equation}
\label{eq-4} \sum_{n=0}^\infty\!{}^{'} f(n) =\int_0^\infty
f(x)\,dx+i\int_0^\infty \frac{f(ix)-f(-ix)}{e^{2\pi x}-1}\;dx{.}
\end{equation}
The first term in the right-hand side of this equation gives the
contribution independent of the temperature, and the net
temperature dependence is produced by the second term in this
formula. Being interested in the low temperature behavior of the
internal energy we substitute into the second term in Eq.\
(\ref{eq-4}) the following  expansion of the function $E_1(z)$
\begin{equation}
\label{eq-5}
E_1(z) =-\gamma -\ln
z- \sum^\infty_{k=1}\frac{(-1)^k z^k}{k\cdot k!},\quad |\arg z |<\pi {,}
\end{equation}
where $\gamma $ is the Euler constant.\cite{AS} The contribution
proportional to $T^3$ is produced by the logarithmic term in the
expansion (\ref{eq-5}). The higher powers of $T$ are generated by
the respective terms in the sum over $k$ in this formula $(t=2\pi a T)$
\begin{equation}
\label{eq-6}
\overline {U}_W(T)=\frac{\Delta n^2}{\pi a}\left ( -\frac{1}{96}+
\frac{\zeta (3)}{4\pi ^2} t^3 -\frac{1}{30}t^4
+\frac{8}{567} t^6
-\frac{8}{1125}t^8+{\cal O}(t^{10})\right )
{.}
\end{equation}
All these terms, safe for $2 \zeta (3) \Delta n^2 a^2T^3$, are
also reproduced by the last term in Eq.\ (3.31) in our paper$^4$
(unfortunately additional factor 4 was missed there)
\[
\frac{\Delta n^2}{8}T\cdot 4\, t^2\int ^2_{\Delta n}\frac {dR}{R}
\frac{\coth (tR)}{\sinh^2 (tR)}{.}
\]
Taking all this into account we arrive at the following low
temperature behavior of the internal Casimir energy of a dilute
dielectric  ball
\begin{equation}
\label{eq-7} U(T)= \frac{\Delta n^2}{\pi a}\left ( \frac{23}{384}
+\frac{\zeta(3)}{4\pi^2}t^3 -\frac{7}{360}t^4
+\frac{22}{2835}t^6 -\frac{46}{7875}t^8 +{\cal O}(t^{10})
 \right ){.}
\end{equation}
The relevant  thermodynamic relations give the following low
temperature expansions for free energy
\begin{equation}
\label{eq-8}
F(T)=\frac{\Delta n^2}{\pi a}\left (
\frac{23}{384}-\frac{\zeta (3)}{8\pi ^2}t^3+\frac{7}{1080}t^4
 -\frac{22}{14175}t^6+\frac{46}{55125}t^8+{\cal O}(t^{10})
\right )
\end{equation}
and for entropy
\begin{equation}
\label{eq-9}
S(T)=-\frac{\partial F}{\partial T}=\Delta n^2
\left (
\frac{3\zeta (3)}{4\pi ^2}t^2-\frac{7}{135}t^3
+\frac{88}{4725}t^5- \frac{736}{55125}t^7+ {\cal O}(t^9)
\right ){.}
\end{equation}

The range of applicability of the  expansions (\ref{eq-7}),
(\ref{eq-8}), and (\ref{eq-9}) can be roughly estimated in the
following way. The curve $S(T)$ defined by Eq.\ (\ref{eq-9})
monotonically goes up when the dimensionless temperature $t =2\pi a
T$ changes from 0 to $t \sim 1.0$. After that  this curve sharply
goes down to the negative values of $S$. It implies  that Eqs.\
(\ref{eq-7}) -- (\ref{eq-9}) can be used in the region $0\leq t <
1.0$. The $T^3$-term in Eqs.\ (\ref{eq-7}) and (\ref{eq-8}) proves to
be principal because it gives the first positive term in the low
temperature expansion for the  entropy (\ref{eq-9}). It is worth
noting, that the exactly the same $T^3$-term, but with opposite sign,
arises in the high temperature asymptotics of free energy in the
problem at hand (see Eq.\ (4.30) in Ref.\cite{BNP}).

For large temperature $T$ we found\cite{NLS}
\begin{equation}
\label{eq-10}
 U(T) \simeq  \frac{\Delta n^2}{8}\, T {,}\;\;
 F(T)  \simeq  -\frac{\Delta n^2}{8}\, T\left [\ln (aT)-c\right ]{,}\;\;
 S(T)\simeq \frac{\Delta n^2}{8}\left [
\ln (aT)+c+1
\right ]
{,}
\end{equation}
where $c$ is a constant\cite{Barton,BNP}
$
c=\ln 4 +\gamma -{7}/{8}\,{.}
$
Analysis of Eqs.\ (3.20) and (3.31) from the paper\cite{NLS} shows that
there are
only exponentially
suppressed corrections to the leading terms (\ref{eq-10}).

The Casimir forces, exerted on the surface of a dielectric ball and
tending to expand it,  have the following low temperature and high
temperature asymptotics
\[
\label{force}
{\cal F} = -\frac{1}{4 \pi a^2}\frac{\partial
F(T)}{\partial a}=\frac{23}{1536} \frac{\Delta n^2}{\pi^2a^4}\left
( 1+\frac{96}{23}\frac{\zeta(3)}{\pi^2}t^3
-\frac{112}{345}t^4 +{\cal O}(t^6)\right )
{,}
\]
\[ {\cal F}\simeq \frac{\Delta n^2}{32 \pi a^3}T, \quad T\to \infty {.}
\]

Summarizing we conclude that now there is a complete agreement
between the results of  calculation of  the Casimir thermodynamic
functions for a dilute dielectric ball carried out in the framework
of two different approaches:  by the mode summation
method\cite{LSN,NLS} and by perturbation theory for quantized
electromagnetic field, when  dielectric ball is considered as a
perturbation in unbounded continuous surroundings.\cite{Barton}

% Authors thank Michael Bordag for excellent organization of the
% Casimir Effect Session at the MG9 Meeting.

\nonumsection{Acknowledgements} \noindent The very pleasant and
friendly atmosphere at the Fifth Workshop on Quantum Field Theory
under the Influence of External Conditions stimulated the elucidation
of many subtle points of this study. V.V.N.\ thanks Professor G.
Barton for fruitful discussions at the Workshop and Professor G.L.\
Klimchitskaya for crucial comment. The partial financial support of
the ISTC (Project No.\ 840) and RFBR (Grant No.\ 00-01-00300) is
acknowledged.

\nonumsection{References}

\eject

\end{document}

\end{document}